\documentclass[useAMS,usenatbib]{mn2e}
\usepackage{graphicx,color}

\usepackage{times,ulem}

\newcommand{\alm}{a_{\ell m}}
\newcommand{\Cl}{C_\ell}
\newcommand{\ClTT}{C_\ell^{TT}}
\newcommand{\ellmax}  {\ell_{\rmn{max}}}
\newcommand{\nside}{\textsc{Nside}}
\newcommand{\muK}{\rmn{\umu K}}
\newcommand{\spice}{\textsc{SpICE}}
\newcommand{\healpix}{\textsc{HEALPix}}

\newcommand*{\satellite}[1]{\textit{#1}}
\newcommand{\COBE}{\satellite{COBE}}
\newcommand{\COBEDMR}{\satellite{COBE-DMR}}
\newcommand{\WMAP}{\satellite{WMAP}}
\newcommand{\planck}{\satellite{Planck}}

\newcommand{\LCDM}{$\Lambda$CDM}

\newcommand*{\unit}[1]{\;\rmn{#1}}
\renewcommand*{\vec}[1]{\bmath{#1}}
\newcommand*{\unitvec}[1]{\vec{\hat{#1}}}

\newcommand{\Shalf}{S_{1/2}}
\newcommand{\STQ}{S^{TQ}}
\newcommand{\sTQ}{s^{TQ}}

\newcommand{\iimag}{\rmn{i}}
\newcommand{\dderiv}{\rmn{d}}

\title[CMB polarization predictions]{Large-Angle CMB Suppression and
  Polarization Predictions}
\author[C.J. Copi, D. Huterer, D.J. Schwarz and G.D. Starkman]
{Craig J. Copi$^{1}$\thanks{E-mail: cjc5@cwru.edu},
  Dragan Huterer$^{2}$\thanks{E-mail: huterer@umich.edu},
  Dominik J. Schwarz$^{3}$\thanks{E-mail: dschwarz@physik.uni-bielefeld.de}
  and 
  Glenn D. Starkman$^{1,4}$\thanks{E-mail: glenn.starkman@case.edu}\\
  $^{1}$CERCA/Department of Physics/ISO, Case Western Reserve University, Cleveland, 
  OH 44106-7079, USA\\
  $^{2}$Department of Physics, University of Michigan, 
  450 Church St, Ann Arbor, MI 48109-1040, USA\\
  $^{3}$Fakult\"at f\"ur Physik, Universit\"at Bielefeld, Postfach 100131, 
  33501 Bielefeld, Germany\\
  $^{4}$Physics Department, Theory Unit, CERN, CH-1211 Gen\`eve 23, Switzerland}

\begin{document}

\date{Accepted xxxx. Received xxxx; in original form xxxx}

\pagerange{\pageref{firstpage}--\pageref{lastpage}} \pubyear{2013}

\maketitle

\label{firstpage}

\begin{abstract}
  The anomalous lack of large angle temperature correlations has been a
  surprising feature of the cosmic microwave background (CMB) since first
  observed by \COBEDMR\ and subsequently confirmed and strengthened by the
  \satellite{Wilkinson Microwave Anisotropy Probe}.  This anomaly may point
  to the need for modifications of the standard model of cosmology or may
  indicate that our Universe is a rare statistical fluctuation within that
  model.  Further observations of the temperature auto-correlation function
  will not elucidate the issue; sufficiently high precision statistical
  observations already exist.  Instead, alternative probes are required.
  In this work we explore the expectations for forthcoming polarization
  observations.  We define a prescription to test the hypothesis that the
  large-angle CMB temperature perturbations in our Universe represent a
  rare statistical fluctuation within the standard cosmological model.
  These tests are based on the temperature-$Q$ Stokes parameter
  correlation.  Unfortunately these tests cannot be expected to be
  definitive.  However, we do show that if this $TQ$-correlation is
  observed to be sufficiently large over an appropriately chosen angular
  range, then the hypothesis can be rejected at a high confidence level.
  We quantify these statements and optimize the statistics we have
  constructed to apply to the anticipated polarization data. We find that
  we can construct a statistic that has a $25$ per cent chance of excluding
  the hypothesis that we live in a rare realization of \LCDM\ at the $99.9$
  per cent confidence level.
\end{abstract}


\section{Introduction}

In the two decades since the \satellite{Cosmic Background Explorer} (\COBE)
first detected the primordial fluctuations in the cosmic microwave
background (CMB) temperature \citep{Wright1992}, and perhaps even more so
in the past decade over which the \satellite{Wilkinson Microwave Anisotropy
  Probe} (\WMAP) has provided ever more accurate full-sky maps of those
fluctuations \citep[see][for example]{WMAP7-cosmology}, the CMB has become
a keystone in the remarkable transition of cosmology from a qualitative to
a precision science.

An important element of the role of the CMB in precision cosmology has been
that the canonical theory of cosmology, inflationary Lambda Cold Dark
Matter (\LCDM), makes clear predictions for the statistical properties of
the spherical harmonic coefficients of the temperature fluctuations,
 \begin{equation}
   \alm \equiv \int Y_{\ell m}^*(\theta,\phi) T(\theta,\phi)\,
   \dderiv(\cos\theta)\,\dderiv\phi,
 \end{equation}
 which are predicted to be statistically isotropic realizations of
 independent Gaussian random variables of zero mean and with variance $\Cl$
 depending only on $\ell$,
\begin{equation}
 \Cl = \frac1{2\ell+1}\sum_{m=-\ell}^\ell \vert \alm\vert^2.
\end{equation}
In modern discussions of the CMB the two-point angular power spectrum,
embodied in these $\Cl$, plays a central role and is the source of the
remarkable precision of the cosmological parameters
\citep{WMAP7-cosmology}.

Before the \COBE\ era, it was the two-point angular correlation function of
the fluctuations,
\begin{equation}
C(\theta) \equiv \overline{T({\unitvec n}_1)T({\unitvec
    n}_2)}\vert_{{\unitvec n}_1\cdot{\unitvec n}_2=\cos\theta},
\end{equation}
rather than the angular power spectrum that was of primary interest to
astronomers.  Statistically, the two-point angular correlation function is
an ensemble average but, in practice, this must be replaced by an average
over pairs of points separated by an angle $\theta$, as denoted by the bar
over the expression.  In fact the \COBE\ differential microwave radiometer
(\COBEDMR) did report $C(\theta)$, though only in their final, four-year
paper \citep{DMR4}.  When extracted from a full-sky map both $C(\theta)$
and the $\Cl$ contain the same information, albeit in different forms.  The
same is true for a function and its Fourier transform; signals are
typically most easily seen in one or the other forms but not both.  In the
case of the CMB, $C(\theta)$ and the $\Cl$ are related by a Legendre
series.  The $\Cl$ most easily show the small angular scale behaviour,
microphysics at last scattering, whereas the $C(\theta)$ most easily shows
the large angular scale behaviour.

As observed by the \COBEDMR, $C(\theta)$ had an unexpected property -- it
was consistent with zero for angular separations between approximately
$60\degr$ and $160\degr$.  This was duly noted at the time but mainly remembered
today as a low quadrupole.  The \WMAP\ team confirmed the
\COBEDMR\ observation of a lack of large-angle correlation with
significantly smaller error bars.  In their initial, one-year release
\cite{WMAP1-cosmology} phrased the anomaly in terms of a statistic
\begin{equation}
  \Shalf \equiv \int_{-1}^{1/2} \left[C(\theta)\right]^2
  \dderiv(\cos\theta).
\end{equation}
In the best-fitting \LCDM\ model the expected value of this statistic is
approximately $50,000\unit{(\muK)^4}$, whereas the observed value is
approximately $8500\unit{(\muK)^4}$ on the full-sky, e.g.\ from the
\WMAP\ independent linear combination (ILC) map,\footnote{The ILC map and
  all data from the \WMAP\ mission used in this work are freely available
  from http://lambda.gsfc.nasa.gov/.}  with a $p$-value of approximately
$0.05$.  Even more striking is that if one considers only the part of the
sky outside a conservative Galaxy cut, then $\Shalf \simeq
1000-1150\unit{(\muK)^4}$ and is only $\sim1300\unit{(\muK)^4}$ in each of
the $V$ and $W$ frequency bands, which are expected to be dominated by the
CMB signal.  The cut-sky $\Shalf$ has a $p$-value of about
$2.5\times10^{-4}$, depending on the precise map \citep{CHSS-WMAP5}.

We have argued that such absence of the two-point angular correlation is
unlikely to result solely from a small quadrupole, or even a small
quadrupole and octopole, and that it instead requires a `conspiracy' among
the first several multipoles \citep{CHSS-WMAP5}.  Such covariance among the
$\Cl$ is likely contrary to the fundamental prediction of the canonical
cosmological model that the $\alm$ underlying the $\Cl$ are independent
Gaussian random variables with variances depending only on $\ell$.  This is
one of a number of large-scale anomalies that suggest that modifications of
the standard model are required on large angular scales \citep[see][and
  references therein for further details]{WMAP7-anomalies,CHSS-review}.

One possible explanation of the absence of large-angle correlations among
the CMB temperature fluctuations is that it is merely a statistical fluke.
In this paper, we explore the consequences of this hypothesis.  In
particular, since the fluctuations in the CMB temperature and in its
polarization arise largely from the same source -- the gravitational
potential -- one might have hoped that a small temperature-temperature ($TT$)
correlation function on large angular scales would predict a similarly
small cross-correlation between CMB temperature and CMB Stokes parameter
$Q$, or in the polarization-polarization ($QQ$) correlation.

Unfortunately, as we shall see, the connection between temperature and
polarization fluctuations is too weak for a general definitive test of the
origin of the vanishing correlation function.  However, we do find that if
the $\Shalf$ is small because of a statistical fluke within
\LCDM\ cosmology, then the cross-correlation between temperature and
polarization is unlikely to be large on large angular scales.  Therefore,
were we to infer a large value of this cross-correlation from future data,
that would be evidence against a statistical fluke as an explanation of the
vanishing $TT$ correlation.

In this paper we provide a prescription to follow in order to test this
hypothesis.  In Section \ref{secn:ConstrainedRealizations}, we describe the
construction of an ensemble of realizations of \LCDM\ that are constrained
to resemble our observed Universe in the properties of their $TT$ angular
power spectrum and full-sky and cut-sky two-point angular correlation
functions.  In Section \ref{secn:Statistics} we construct statistics that,
like $\Shalf$ for the $TT$ correlations, can be used to quantify the
smallness of the correlations between temperature and polarization
fluctuations.  Section \ref{secn:Results} contains a discussion of the
results of applying the temperature-polarization cross-correlation
statistics to the ensemble of constrained realizations and looks forward to
what might be learned by applying them to future polarization data.  Given
the constructed realizations the value of the statistics are fixed and the
optimal application of them to polarization data can be determined as
discussed in this section.  This optimization's is \textit{independent} of
polarization observations.  Finally, Section \ref{secn:Conclusions}
contains the conclusions.

\section{Constrained realizations}
\label{secn:ConstrainedRealizations}

To study the signature of the lack of large-angle correlations in the
\WMAP\ temperature data on upcoming polarization measurements, such as from
\planck, we require realizations of \LCDM\ consistent with these
large-angle results.  For this purpose, we have generated $300\,000$ such
realizations as follows.
\begin{enumerate}
\item
  In the standard, \LCDM\ model our Universe is a realization from an
  ensemble, the width of which (the cosmic variance) for low-$\ell$
  information is quite large. However, once measured our realization can
  and has been precisely determined.  In the work reported here this
  information is given in the \WMAP\ reported $\ClTT$\@.  We are interested
  in producing realizations of \textit{our} Universe as represented by the
  \WMAP\ observations, \textit{not} realizations of the full \LCDM\ model.
  For this purpose we treat the observational errors in the \WMAP\ reported
  $\ClTT$ as Gaussian distributed and generate realizations accordingly.
  Thus we generate random $\ClTT$ from Gaussian distributions centred on
  the \WMAP-reported values.  This produces a power spectrum consistent
  with that reported by \WMAP\@.  Again it is important to stress that this
  is a power spectrum consistent with the observation of our particular
  realization of the Universe as measured by \WMAP, not a general
  realization of the best-fitting \LCDM\ model.  For this reason, cosmic
  variance is not relevant nor do we generate $\Cl$ from a $\chi^2$
  distribution.  It is also true that on a partial sky the $\Cl$ are
  slightly correlated.  To correct for this we actually use the Fisher
  matrix from the \WMAP\ likelihood code \textit{without} the contribution
  from cosmic variance in drawing these $\ClTT$.  In practice this is a
  small correction but has been included for completeness.
\item 
  The $\ClTT$ generated in the previous step contain the statistical
  information about the power in each mode consistent with the
  \WMAP\ observations.  For further analysis we need a map, not just the
  power spectrum.  A map is a particular realization of this power
  spectrum. As in \LCDM\ we assume that the modes in the map have random
  phases.  In practice, this means we choose the $\alm^T$ randomly on the
  $2\ell$-sphere such that
  \begin{equation}
    \frac{1}{2\ell+1} \sum_{\ell=-m}^m \left|\alm^T\right|^2 = \ClTT,
  \end{equation}
  where the $\ClTT$ in this expression are \textit{exactly} the values
  generated from the previous step.  By this construction the resulting sky
  realization is guaranteed to have a $\Shalf$ consistent with the small
  value in the full-sky \WMAP\ ILC map.  For each set of $\ClTT$ from the
  previous step we generate a single complete sky realization, i.e.\ a
  single map.
\item To further be consistent with \WMAP\ observations the $\Shalf$ on the
  cut sky must also be small.  For our realizations we require
  $\Shalf^{\rmn{cut}} \le 1292.6\unit{(\muK)^4}$, the value from the
  \WMAP\ seven-year, KQ75y7 masked ILC map.  This $\Shalf$ value is
  calculated for a realization by first constructing a map at \nside=64
  from the $\alm^T$ generated above.  The pseudo-$\Cl$ are extracted from
  this map based on the region outside the KQ75y7 mask using
  \spice\ \citep{polspice}.  Finally $\Shalf$ is calculated using these
  $\Cl$ up to $\ellmax=100$.
  \item For temperature realizations that satisfy the cut sky constraint we
    also generate realizations of the $\alm^E$.  See
    Appendix~\ref{app:almE-realization} for a review of the process.
\end{enumerate}
The construction of a set of constrained realizations is the basis for the
prescription we are describing.  It will next provide predictions for the
expectations from the observations of the CMB polarization.

\section{Statistics}
\label{secn:Statistics}

For the temperature auto-correlation, the $\Shalf$ statistic was defined
\textit{a posteori} \citep{WMAP1-cosmology} to be
\begin{equation}
\label{eqn:Shalf_TT}
  \Shalf \equiv \int_{-1}^{1/2} \left[ C^{TT}(\theta) \right]^2
  \dderiv(\cos\theta).
\end{equation}
Inspired by this we define a comparable statistic for $C^{TQ}(\theta)$, the
two-point angular correlation function between fluctuations in the
temperature and the Stokes parameter $Q$.

Observable properties of photons can be characterized by the Stokes
parameters.  For the CMB the relevant quantities are the intensity,
conventionally represented by the temperature, $T$, and the linear
polarization given by the $Q$ and $U$ parameters.  For the CMB the circular
polarization, represented by the $V$ Stokes parameter, is expected to be
zero and not considered further.  When working in real space the natural
correlations to construct are among these observables, $T$, $Q$, and $U$.
These correlations are constructed such that they only depend on the
angular separation along the great circle connecting each pairs of point on
the sky and are thus rotationally invariant despite the fact that the
definition of $Q$ and $U$ depend on the choice of coordinate axes
\citep{Kamionkowski1997}.  When working in harmonic space, it is natural to
decompose the polarization into `gradient' and `curl' modes
\citep{Kamionkowski1997} alternatively called $E$ and $B$ modes
\citep{Zaldarriaga1997}, which are similarly rotationally invariant
quantities.  These latter names will be used throughout.  Thus, in real
space we will work with the $TQ$ two-point angular correlation function,
$C^{TQ}(\theta)$, which may be written in terms of the two-point angular
power spectrum coefficients, $\Cl^{TE}$.

\begin{figure}
  \includegraphics[width=0.5\textwidth]{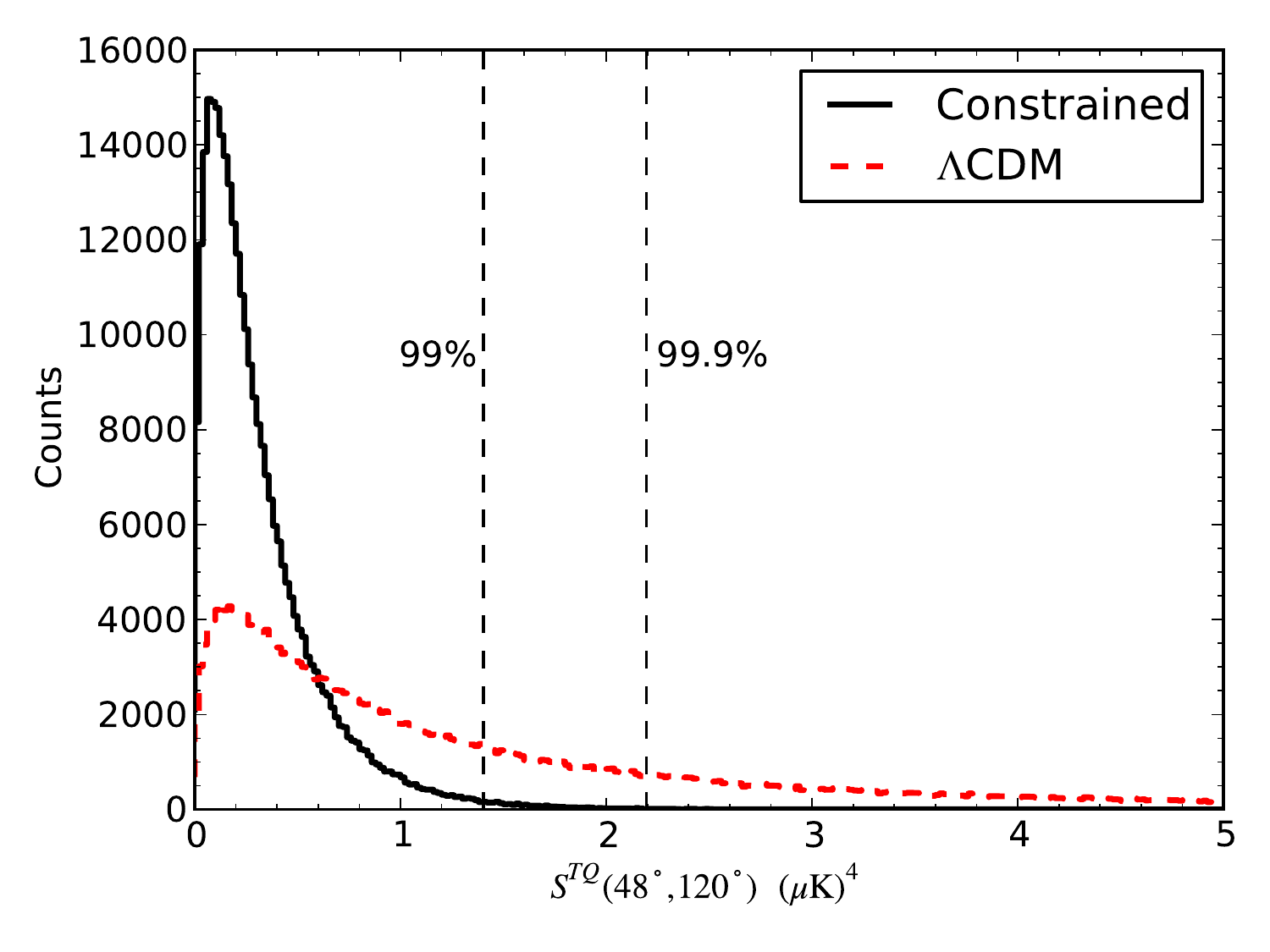}
  \caption{Example histogram of the $\STQ$ statistic, defined in
    Eq.~(\ref{eq:STQ}), for constrained (solid, black line) and
    \LCDM\ (dashed, red line) realizations.  We note that the constrained
    realizations are more sharply peaked at low $\STQ$ than \LCDM\ which,
    though peaked at approximately the same value, has a long tail.  The
    dashed, vertical lines represent the values of $\STQ$ for which $99$
    per cent and $99.9$ per cent, respectively, of the constrained
    realizations have smaller values.}
  \label{fig:STQ-histogram}
\end{figure}

\subsection{$\bmath{\STQ}$ Statistic}

In the case of polarization \textit{a priori} the optimal range over which
to integrate the correlation function is unknown and will be explored below
so we define the general statistic
\begin{equation}
  \STQ(\theta_1,\theta_2) \equiv \int_{\cos\theta_2}^{\cos\theta_1}
  \left[ C^{TQ}(\theta) \right]^2 \dderiv(\cos\theta).
  \label{eq:STQ}
\end{equation}
As with $\Shalf$ we may calculate this easily in terms of the power
spectrum coefficients, $\Cl^{TE}$. 
Using \citep{Kamionkowski1997}
\begin{equation}
  C^{TQ}(\theta) = \sum_{\ell=2}^\infty \frac{2\ell+1}{4\upi}
  \sqrt{\frac{(\ell-2)!}{(\ell+2)!}} \Cl^{TE} P_\ell^2(\cos\theta)
\end{equation}
we may show that
\begin{equation}
  \STQ(\theta_1,\theta_2) = \sum_{\ell,\ell'} C^{TE}_{\ell}
  I^{TQ}_{\ell,\ell'}(\theta_1,\theta_2) C^{TE}_{\ell'},
\end{equation}
where $I^{TQ}_{\ell,\ell'} (\theta_1,\theta_2)$ are components of a known
matrix calculated in Appendix~\ref{app:statistics}.  A histogram of the
$\STQ$ statistic for a particular choice of $\theta_1$ and $\theta_2$ is
shown in Fig.~\ref{fig:STQ-histogram} comparing the constrained
realizations to \LCDM.

\begin{figure}
  \includegraphics[width=0.5\textwidth]{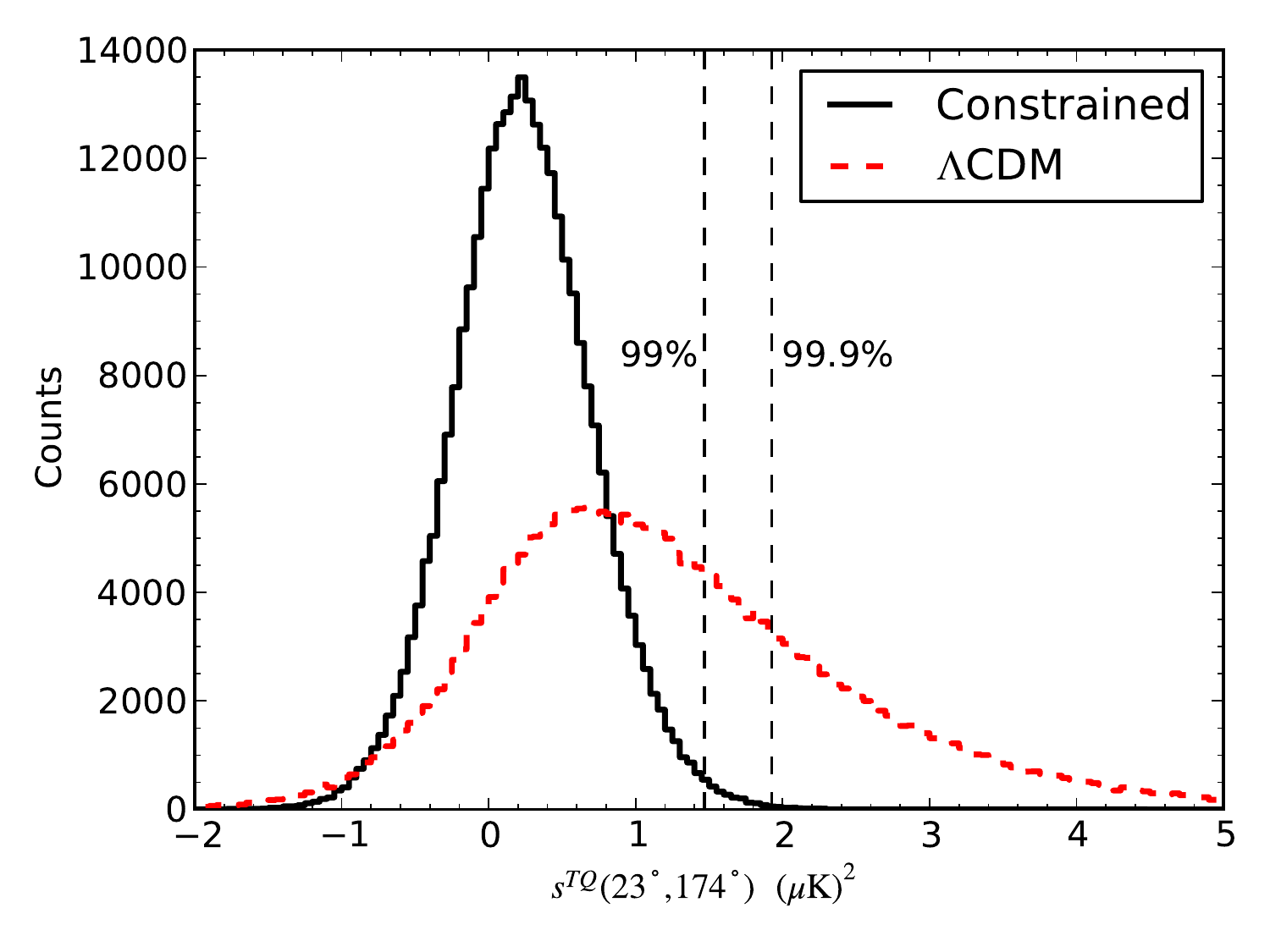}
  \caption{Example histogram of the $\sTQ$ statistic, defined in
    Eq.~(\ref{eq:sTQ}), for constrained (solid, black line) and
    \LCDM\ (dashed, red line) realizations.  We note that the constrained
    realizations are more sharply peaked near zero than \LCDM\ which,
    though also peaked near zero, has a long tail particularly to large,
    positive values.  The dashed, vertical lines represent the values of
    $\sTQ$ for which $99$ per cent and $99.9$ per cent, respectively, of
    the constrained realizations have smaller values.}
  \label{fig:sTQ-histogram}
\end{figure}

\subsection{$\bmath{\sTQ}$ Statistic}

Motivated solely by its simplicity and ease of computation we also define a
new statistic which is linear, rather than quadratic, in the $TQ$
correlation function
\begin{equation}
  \sTQ(\theta_1,\theta_2) \equiv \int_{\cos\theta_2}^{\cos\theta_1}
  C^{TQ}(\theta) \, \dderiv(\cos\theta).
  \label{eq:sTQ}
\end{equation}
As with $\STQ(\theta_1,\theta_2)$ we may calculate this easily in terms of the
$C^{TE}_\ell$,
\begin{equation}
  \sTQ(\theta_1,\theta_2) = \sum_{\ell=2}^\infty   C^{TE}_{\ell}
  i^{TQ}_{\ell}(\theta_1,\theta_2),
\end{equation}
where $i^{TQ}_\ell (\theta_1,\theta_2)$ are the components of a known
vector calculated in Appendix~\ref{app:statistics}.  A histogram of the
$\sTQ$ statistic for a particular choice of $\theta_1$ and $\theta_2$ is
shown in Fig.~\ref{fig:sTQ-histogram} comparing the constrained
realizations to \LCDM.

\section{Results}
\label{secn:Results}

The $\STQ$ and $\sTQ$ statistics defined above have been calculated for the
constrained realizations discussed in
Section~\ref{secn:ConstrainedRealizations} and for a comparable number of
realizations of \LCDM\@.  In both cases these have been calculated from
maps produced at \nside=64.  Data from the temperature map outside the
KQ75y7 mask and from the seven-year polarization analysis mask, both
provided by \WMAP, have been used. As shown in
Figs.~\ref{fig:STQ-histogram} and \ref{fig:sTQ-histogram}, the constrained
and \LCDM\ realizations have most likely values for the $\STQ$ and $\sTQ$
statistics at nearly the same value.  However, as we also see
\LCDM\ predicts much broader distributions for the two statistics.  In
particular, there is a significant probability of producing values larger
than the constrained realizations.  This provides a means of testing the
hypothesis that our Universe is just a rare realization of \LCDM\@.  This
results in a simple but not definitive test.

Consider the case of the $\STQ$ statistic as represented in
Fig.~\ref{fig:STQ-histogram}.  For the constrained realizations $99$ per
cent of them have $\STQ(48\degr,120\degr)\le 1.403\unit{(\muK)^4}$ and
$99.9$ per cent have $\STQ(48\degr,120\degr)\le 2.195\unit{(\muK)^4}$.
Unconstrained \LCDM\ (with the best-fitting values of cosmological
parameters) randomly generates realizations with values larger than these
$38.6$ per cent and $25.6$ per cent of the time, respectively.  If
observations of the polarization show our Universe to have a
$\STQ(48\degr,120\degr)$ value larger than these values, then we can reject
the random \LCDM\ realization hypothesis at the appropriate confidence
level.  Alternatively, if the $\STQ(48\degr,120\degr)$ value is smaller,
then no definitive statement can be made; the polarization fluctuations
would be consistent with the hypothesis that we live in a rare
\LCDM\ realization but do nothing to advance that hypothesis.  This is the
main point of the paper.

Similar statements may be made about the $\sTQ$ statistic shown in
Fig.~\ref{fig:sTQ-histogram}.  It provides similar information as $\STQ$.

We still have freedom to choose the optimal range of angles over which to
evaluate the statistics.  We define optimal to mean the maximum
discriminatory power between the distribution of the statistic in the
constrained versus \LCDM\ realizations.  For a given per cent cutoff from
the constrained realizations we wish to find the angle range
$[\theta_1,\theta_2]$ that has the maximum fraction of \LCDM\ realizations
above this value.

\begin{figure}
  \includegraphics[width=0.5\textwidth]{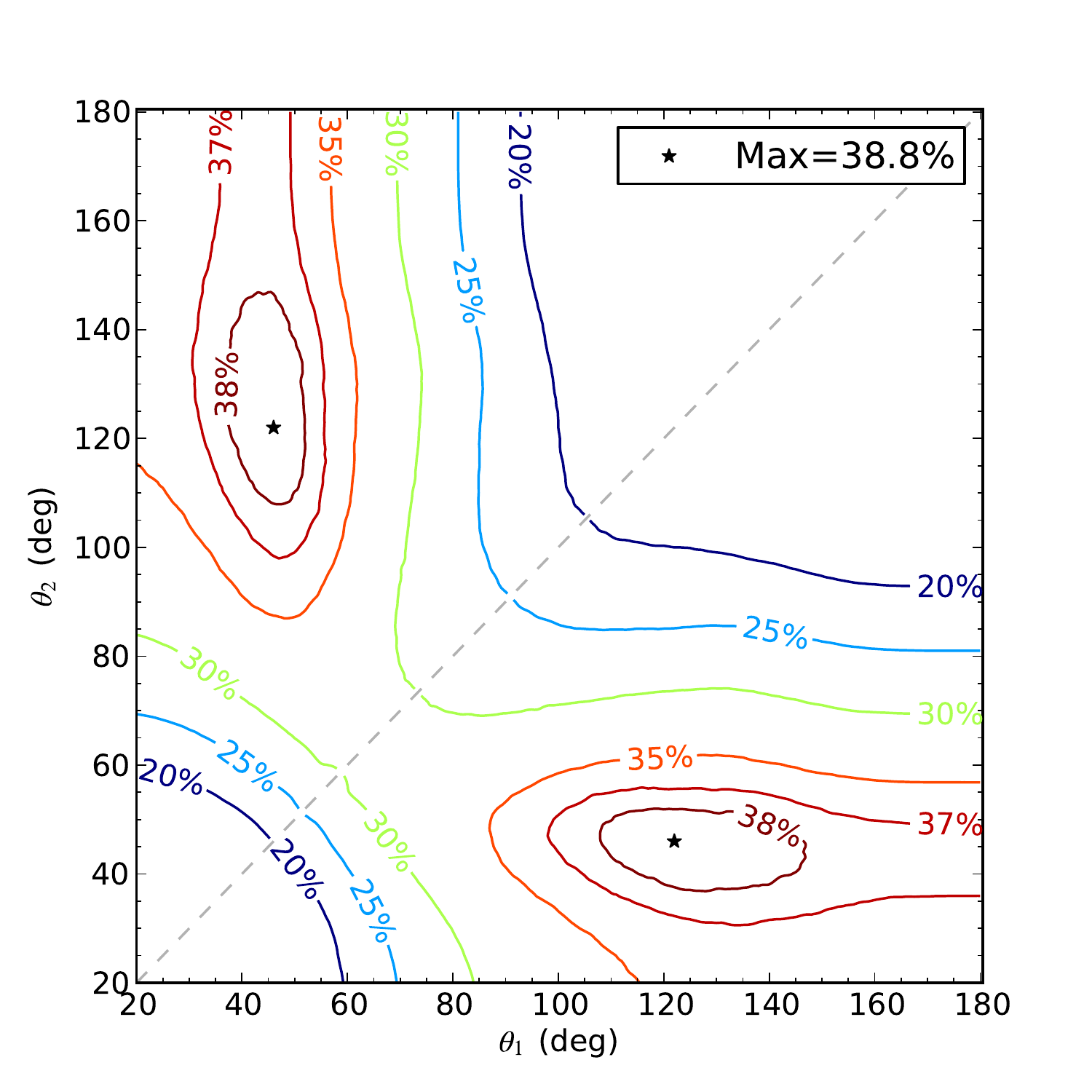}
  \caption{Contours for the fraction of \LCDM\ realizations above the $99$
    per cent value of the constrained realizations from the $\STQ$
    statistic, defined in Eq.~(\ref{eq:STQ}). In the optimal case $38.8$
    per cent of the \LCDM\ realizations have a larger value.  In this
    figure the results are not defined along the diagonal (black, dashed
    line) and have been made symmetric about it by taking the absolute
    value of the statistic.}
  \label{fig:STQ-contours-masked-99}
\end{figure}
\begin{figure}
  \includegraphics[width=0.5\textwidth]{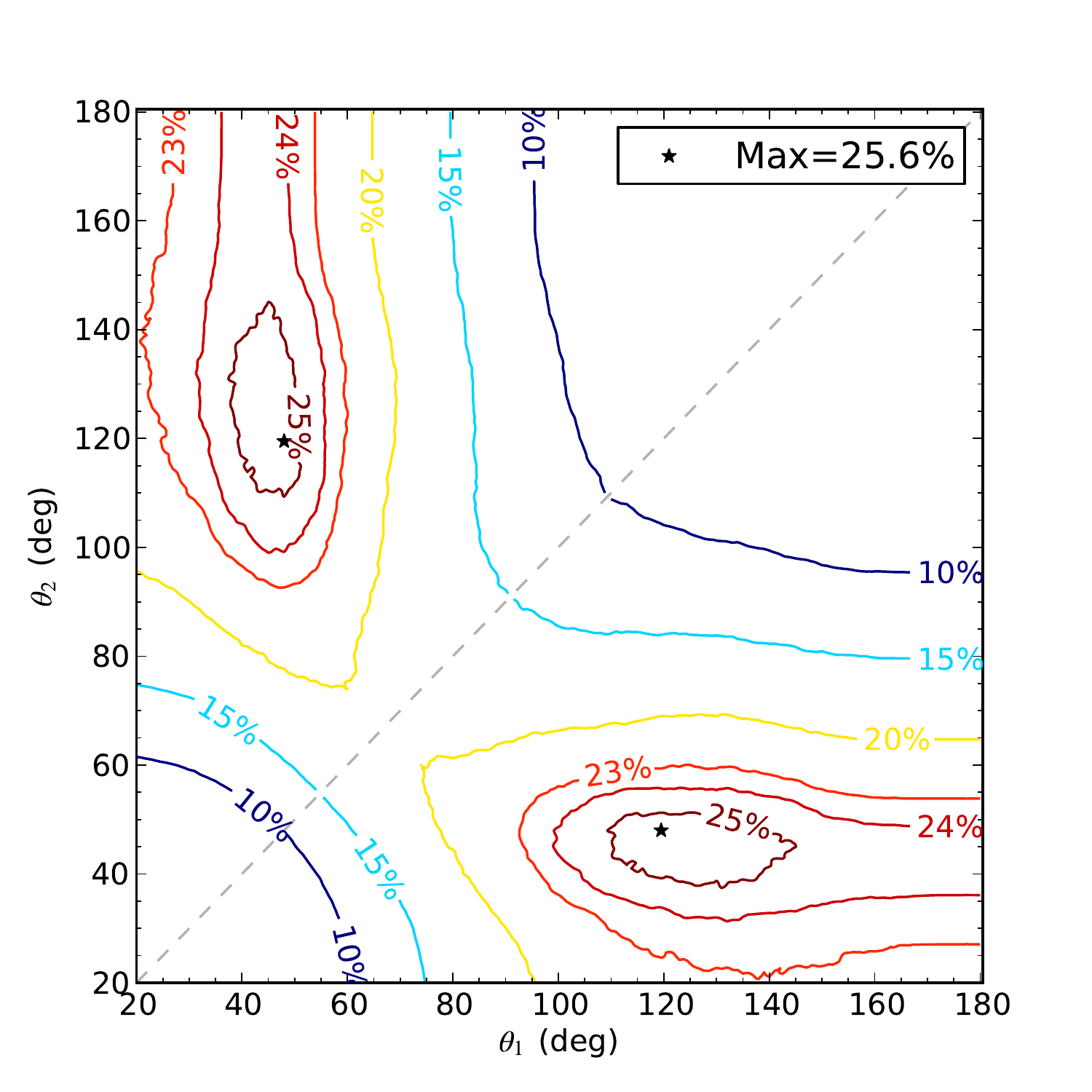}
  \caption{Same as Fig.~\ref{fig:STQ-contours-masked-99} now for the
    $99.9$ per cent case.  Here in the optimal case $25.6$ per cent of the
    \LCDM\ realizations have a larger value and the full histograms are
    shown in Fig.~\ref{fig:STQ-histogram}.}
  \label{fig:STQ-contours-masked-99.9}
\end{figure}

\begin{figure}
  \includegraphics[width=0.5\textwidth]{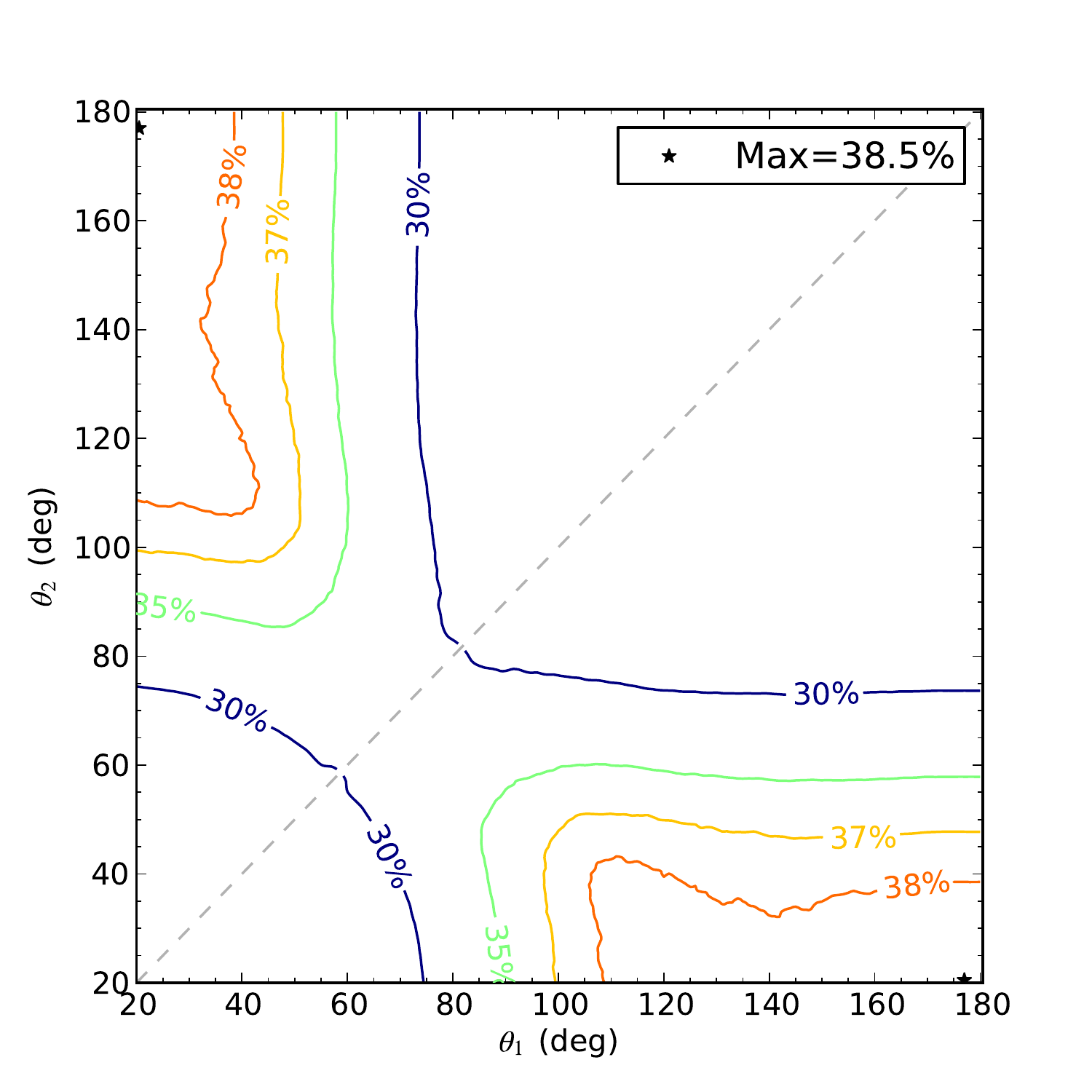}
  \caption{Same as Fig.~\ref{fig:STQ-contours-masked-99} but now for the
    $\sTQ$ statistic, defined in Eq.~(\ref{eq:sTQ}).  Here in the optimal
    case $38.5$ per cent of the \LCDM\ realizations have a larger value at
    the $99$ per cent level.}
  \label{fig:sTQ-contours-masked-99}
\end{figure}

\begin{figure}
  \includegraphics[width=0.5\textwidth]{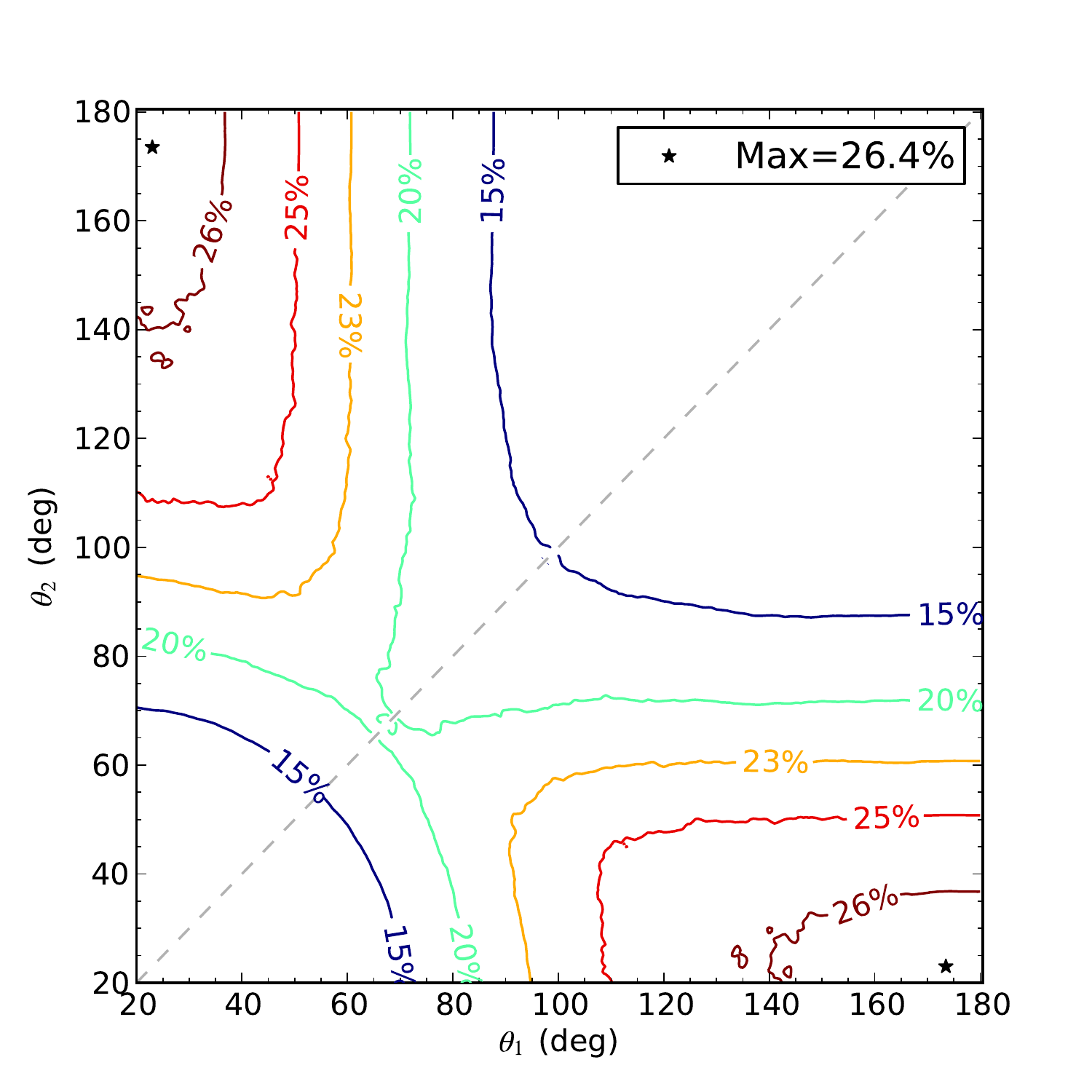}
  \caption{Same as Fig.~\ref{fig:sTQ-contours-masked-99} now for the $99.9$
    per cent case.  Here in the optimal case $26.4$ per cent of the
    \LCDM\ realizations have a larger value and the full histograms are
    shown in Fig.~\ref{fig:sTQ-histogram}.}
  \label{fig:sTQ-contours-masked-99.9}
\end{figure}

The results of such a study are shown in
Figs.~\ref{fig:STQ-contours-masked-99} and
\ref{fig:STQ-contours-masked-99.9} for the $\STQ(\theta_1,\theta_2)$
statistic and in Figs.~\ref{fig:sTQ-contours-masked-99} and
\ref{fig:sTQ-contours-masked-99.9} for the $\sTQ(\theta_1,\theta_2)$
statistic.  The statistics are non-zero only up to the diagonal
$\theta_1=\theta_2$, shown as the dashed, black line in the figures, but
not along it.  For this reason, the contours are truncated at the diagonal.
They have been made symmetric in $\theta_1$ and $\theta_2$ (by taking
$\vert\STQ\vert$ and $\vert\sTQ\vert$) so the results are shown as
identical when reflected through the diagonal.  The optimal ranges and
fractions of \LCDM\ realizations are listed in
Table~\ref{tab:optimal-ranges}.  We note that the optimal surfaces
represented by the contours seen in the figures are relatively broad, at
least in one direction.  Due to this the values of $\theta_1$ and
$\theta_2$ in a neighbourhood of those listed in
Table~\ref{tab:optimal-ranges} can be employed with nearly the same
efficacy.

\begin{table}
  \caption{Optimal angle ranges for the $\STQ$ statistic~(\ref{eq:STQ}) and
    $\sTQ$ statistic~(\ref{eq:sTQ}).  The optimal ranges are determined by
    finding the maximum fraction of \LCDM\ realizations with the
    appropriate statistic above the $99$ per cent or $99.9$ per cent level
    of the constrained realizations.  The histograms for the optimal $99.9$
    per cent ranges are shown in Figs.~\ref{fig:STQ-histogram} and
    \ref{fig:sTQ-histogram}.  The full contours are shown in
    Figs.~\ref{fig:STQ-contours-masked-99}--\ref{fig:sTQ-contours-masked-99.9}.}
  \label{tab:optimal-ranges}
  \begin{tabular}{cccccl} \hline
    \textbf{Statistic} & \textbf{C.L.} & $\bmath{\theta_1}$ &
    $\bmath{\theta_2}$ & \textbf{Fraction} \\
    & \textbf{(per cent)} & \textbf{(deg)} & \textbf{(deg)} &
    \textbf{(per cent)} \\ \hline
    $\STQ$ & $99\hphantom{.0}$ & $46$ & $122$ & $38.8$ \\
    & $99.9$ & $48$ & $120$ & $25.6$ \\
    \hline
    $\sTQ$ & $99\hphantom{.0}$ & $20$ & $177$ & $38.5$ \\
    & $99.9$ & $23$ & $174$ & $26.4$ \\
    \hline
  \end{tabular}
\end{table}

\section{Conclusions}
\label{secn:Conclusions}

The absence of two-point angular correlations on large angular scales in
the CMB temperature data is, by now, well established.  It was first
measured by \COBEDMR, but became more significant in the \WMAP\ temperature
maps.  This absence of correlation is difficult to accommodate within the
standard cosmological model, especially since it seems to imply covariance
among low-$\ell$ multipoles of the CMB\@.  A simple explanation that has
been proffered is that we just live in a rare realization of \LCDM\ that
happens to have a lack of large-angle $TT$ correlations.  If so, one might
hope that constraining \LCDM\ realizations to have low $TT$ correlations at
large angles, would have observable consequences for other correlation
functions, such as $TQ$. These would be the basis for an observational test
of this statistical fluke hypothesis.

In this paper, we have discussed a prescription to follow in order to test
this hypothesis of our Universe being a statistical fluke.  The
prescription may be simply stated as follows.
\begin{enumerate}
\item Construct realizations of our Universe consistent with the observed
  temperature fluctuations.  This means construct sets of $\alm^T$ and
  $\alm^E$ consistent with the observed $\ClTT$ and full and cut sky
  $\Shalf$ as discussed in Section~\ref{secn:ConstrainedRealizations}.
\item Apply the $\STQ$ and $\sTQ$ statistics as defined in
  Section~\ref{secn:Statistics} to these constrained realizations.
\item Also apply the $\STQ$ and $\sTQ$ statistics to a comparable number of
  best-fitting \LCDM\ realizations and use these to find the optimal ranges
  $[\theta_1,\theta_2]$ for each statistic.  Optimal here means that the
  maximum fraction of \LCDM\ realizations fall above the value at some
  confidence level in the constrained realization, e.g.\ the $99$ per cent
  or $99.9$ per cent level.
\item Given the optimal ranges from the previous step now apply these
  particular cases to the observed polarization signal.  If the
  observations produce values for these statistics larger than that
  expected from the constrained realizations, then the statistical fluke
  hypothesis can be rejected at the appropriate confidence level.
  Alternatively, if the values are smaller, then the hypothesis remains
  consistent but unproven.
\end{enumerate}
We further note that the optimization's in this prescription is independent
of the polarization observations, or, in fact, whether the polarization has
been observed or not.

Our work is, in spirit, related to \cite{Dvorkin2008}. While they consider
observables in the polarization signal for `models' of three dimensional
primordial power modulation that might explain the breaking of statistical
isotropy in the temperature field, we predict the polarization statistics
starting directly from realizations of \LCDM\ models that are constrained to
show the suppressed $TT$ correlation at large angular scales.

In the work reported here we have generated realizations and performed the
optimization's based on the \WMAP\ seven-year data release.  The prescription
could be applied to the \WMAP\ nine-year data release and the results are
not expected to differ significantly.  We have also said nothing about
applying the statistics to the \WMAP\ reported polarization observations.
Unfortunately, the signal-to-noise ratio in the polarization observations
is not yet sufficient to make meaningful statements.  To see this, using
the \WMAP\ nine-year reported $\Cl^{TE}$ \citep{WMAP9-cosmology}, we find
for the optimal ranges
\begin{equation}
  \STQ_{\WMAP} (48\degr,120\degr) = (1.0\pm0.8)\unit{(\muK)^4}
\end{equation}
and
\begin{equation}
  \sTQ_{\WMAP} (23\degr,174\degr) = (0.8\pm0.8)\unit{(\muK)^2}.
\end{equation}
Here the error bars are crude estimates assuming that the reported
$\Cl^{TE}$ are statistically independent and the noise is Gaussian.  These
assumptions are not justified and a more careful assessment could be
performed using the Fisher matrix.  However, given the large estimated
errors such an assessment is not warranted.

We have shown that the prescription described in this work is not a
definitive test of the statistical fluke hypothesis for our Universe.
Nevertheless, by carefully optimizing the statistical measure of
large-angle $TQ$ correlations, we were able to demonstrate that once good
$TQ$ correlation data are available there is a reasonable probability (over
$25$ per cent) to reject the statistical fluke hypothesis at the $99.9$ per
cent confidence level.  \WMAP\ data are not up to this task; however,
\planck\ data should be.

\section*{Acknowledgements}

GDS and CJC are supported by a grant from the US Department of Energy to
the Particle Astrophysics Theory Group at CWRU\@.  DH has been supported by
the DOE, NASA, and NSF\@.  DJS is supported by the DFG grant RTG 1620
`Models of gravity'.  GDS thanks the Theory Unit at CERN for their
hospitality.  This work made extensive use of the \healpix{}
package~\citep{healpix}.  The numerical simulations were performed on the
facilities provided by the Case ITS High Performance Computing Cluster.

\bibliographystyle{mn2e_new}
\bibliography{polarization_predictions}

\appendix

\section{Polarization Gaussian Random Realizations}
\label{app:almE-realization}

The generation of correlated Gaussian random variables is a well known
topic.  For use in the CMB this is implemented in
\healpix\ \citep{healpix}, for example.  Here we review the details
relevant for the generation of our constrained realizations.

In \LCDM\ the temperature and $E$-type polarization are correlated as
encoded in the power spectrum coefficients $\ClTT$, $\Cl^{TE}$, and
$\Cl^{EE}$ from the best-fitting \LCDM\ model.  Working in the real spherical
harmonic basis we may generate the spherical harmonic coefficients as
\begin{eqnarray}
  a_j^T & = & \sqrt{\ClTT} \zeta_1,
  \label{eq:ajT} \\
  a_j^E & = & \frac{\Cl^{TE}}{\sqrt{\ClTT}} \zeta_1 +
  \sqrt{\Cl^{EE}-\frac{(\Cl^{TE})^2}{\ClTT}} \zeta_2,
  \label{eq:ajE}
\end{eqnarray}
where $\zeta_1$ and $\zeta_2$ are Gaussian random variables drawn from a
distribution with zero mean and unit variance and the index $j$ refers to
the pair of indices $(\ell,m)$.  To be precise, $j$ takes the values $0$ to
$2\ell$ and the complex coefficients are constructed as
\begin{equation}
  \alm^T = \left\{ \begin{array}{cl}
    a_0^T, & m = 0 \\
    \frac1{\sqrt{2}} \left( a_{2m-1}^T + \iimag\, a_{2m}^T \right), & m>0
  \end{array} \right. .
\end{equation}

For our purposes we need to generate constrained realizations of \LCDM\ so
the above procedure must be modified.  The steps discussed in
Section~\ref{secn:ConstrainedRealizations} lead to the generation of
constrained $\alm^T$.  In other words, we have determined $a_j^T$ which
by Eq.~(\ref{eq:ajT}) means we have also determined $\zeta_1$.  That is,
instead of choosing $\zeta_1$ as a Gaussian random variable we have used
observational constraints to determine its value and find it by inverting
that equation.  Since the temperature and $E$-type polarization are
correlated this constrained temperature realization affects $a_j^E$\@.  The
real and imaginary components of $\alm^E$ may now be generated
from~(\ref{eq:ajE}) as
\begin{equation}
  a_j^E = \frac{\Cl^{TE}}{\ClTT} a_j^T +
  \sqrt{\Cl^{EE}-\frac{(\Cl^{TE})^2}{\ClTT}} \zeta_2,
\end{equation}
where $\zeta_2$ is still to be chosen as a Gaussian random variable.

\section{Derivation of Statistics Formulas}
\label{app:statistics}

\subsection{$\bmath{\STQ(\theta_1,\theta_2)}$}

We wish to evaluate $\STQ(\theta_1,\theta_2)$ as discussed in the
text~(\ref{eq:STQ}).  Consider the simpler case
\begin{eqnarray}
  \STQ(x) &\equiv& \int_{-1}^x \left[ C^{TQ}(\theta) \right]^2
  \dderiv(\cos\theta)  \\
  & = &\sum_{\ell,\ell'} \frac{(2\ell+1)(2\ell'+1)}{(4\upi)^2} \sqrt{
    \frac{(\ell-2)!(\ell'-2)!}{(\ell+2)!(\ell'+2)!}} C^{TE}_\ell
  C^{TE}_{\ell'} \nonumber \\
  & & {} \times \int_{-1}^x P_\ell^2(\cos\theta) P_{\ell'}^2(\cos\theta)
  \, \dderiv(\cos\theta). \nonumber
\end{eqnarray}
To evaluate this expression we need to perform the integral of two associated
Legendre functions, $P_{\ell}^{m}$, of order $m=2$,
\begin{equation}
  \tilde I^{TQ}_{\ell,\ell'}(x) \equiv \int_{-1}^x P_\ell^2(x) P_{\ell'}^2(x) \,\dderiv x.
\end{equation}
Note that this integral is only defined for $\ell,\ell'\ge 2$.

For $\ell\ne \ell'$ we may proceed by directly integrating the associated
Legendre differential equation to find
\begin{eqnarray}
  \tilde I^{TQ}_{\ell,\ell'}(x) & = & \frac{1-x^2}{\ell(\ell+1)-\ell'(\ell'+1)} \\
  && {} \times \left[ P_\ell^2(x)
    \frac{\dderiv P_{\ell'}^2(x)}{\dderiv x} - P_{\ell'}^2(x)
    \frac{\dderiv P_\ell^2(x)}{\dderiv x} \right].
  \nonumber
\end{eqnarray}

For $\ell=\ell'$ more care is required.  Starting from the Rodriguez formula
\begin{equation}
  P_\ell^2(x) = (1-x^2)\frac{\dderiv^2 P_\ell(x)}{\dderiv x^2}
  \label{eq:Pn2-rodriguez}
\end{equation}
and the recursion relation
\begin{equation}
  x\frac{\dderiv P_\ell(x)}{\dderiv x}  = \frac{\dderiv P_{\ell-1}(x)}{\dderiv
    x} + \ell P_\ell(x)
\end{equation}
we can show that
\begin{equation}
  \tilde I^{TQ}_{\ell,\ell}(x) = 4 J_{\ell-1}^{(2)}(x) - 4\ell (\ell-1) J_\ell^{(1)}(x) +
    \ell^2(\ell-1)^2 \tilde I_{\ell,\ell}(x).
\end{equation}
Here $\tilde I_{\ell,\ell}(x)$ is the equivalent integral over Legendre
polynomials encountered in the definition of $\Shalf$; see appendix A of
\cite{CHSS-WMAP5} for details.  The remaining quantities, $J_\ell^{(1)}(x)$ and
$J_\ell^{(2)}(x)$, are calculated through integration by parts and use of the
recursion relation to find
\begin{eqnarray}
  J_\ell^{(1)}(x) & = & P_{\ell-1}(x) P_\ell(x) \\
  & & {} + \frac12 \left\{ 1 - x \left[ P_{\ell-1}(x)
    \right]^2 - (2\ell-1) \tilde I_{\ell-1,\ell-1}(x) \right\}, \nonumber
\end{eqnarray}
and
\begin{equation}
  J_\ell^{(2)}(x) = J_{\ell-1}^{(2)} (x) + \ell \left[ P_{\ell-1}(x) P_\ell(x) + 1 \right].
\end{equation}
Note that $J_\ell^{(2)}(x)$ is defined recursively.  We can directly show
that $J_0^{(2)}(x)=0$.

With these expressions for the integrals we may write
\begin{equation}
  \STQ(\theta_1,\theta_2) = \sum_{\ell,\ell'} C^{TE}_\ell
  I^{TQ}_{\ell,\ell'}(\theta_1,\theta_2) C^{TE}_{\ell'},
\end{equation}
where
\begin{eqnarray}
  I^{TQ}_{\ell,\ell'}(\theta_1,\theta_2) & = &
  \frac{(2\ell+1)(2\ell'+1)}{(4\upi)^2}
  \sqrt{\frac{(\ell-2)!(\ell'-2)!}{(\ell+2)!(\ell'+2)!}} \\
  & & {} \times
  \left[ \tilde I^{TQ}_{\ell,\ell'}(\cos\theta_1) 
    - \tilde I^{TQ}_{\ell,\ell'}(\cos\theta_2) \right].
  \nonumber
\end{eqnarray}
This matrix may be precomputed for rapid evaluation of
$\STQ(\theta_1,\theta_2)$.

\subsection{$\bmath{\sTQ(\theta_1,\theta_2)}$}

We wish to evaluate $\sTQ(\theta_1,\theta_2)$ as discussed in the
text~(\ref{eq:sTQ}).  We proceed as above and consider the simpler case
\begin{eqnarray}
  \sTQ(x) &\equiv& \int_{-1}^x C^{TQ}(\theta) \, \dderiv(\cos\theta)  \\
  & = &\sum_{\ell=2}^\infty \frac{(2\ell+1)}{4\upi} \sqrt{
    \frac{(\ell-2)!}{(\ell+2)!}} C^{TE}_\ell \nonumber \\
  && {} \times
  \int_{-1}^x P_\ell^2(\cos\theta) \, \dderiv(\cos\theta). \nonumber
\end{eqnarray}
To evaluate this expression we need to perform the integral
\begin{equation}
  \tilde\imath^{TQ}_\ell(x) \equiv \int_{-1}^x P_\ell^2(x)\, \dderiv x.
\end{equation}
This integral is straight-forward to evaluate starting from the Rodriguez
formula~(\ref{eq:Pn2-rodriguez}) and integrating by parts to find
\begin{eqnarray}
  \tilde\imath^{TQ}_\ell(x) & = &
  \ell P_{\ell-1}(x) - (\ell-2) x P_\ell(x) + 2 (-1)^\ell \\
  && {} - \frac{2}{2\ell+1} \left[ P_{\ell+1}(x) - P_{\ell-1}(x) \right]. \nonumber
\end{eqnarray}
With this we may write
\begin{equation}
  \sTQ(\theta_1,\theta_2) = \sum_{\ell=2}^\infty   C^{TE}_\ell
  i^{TQ}_\ell(\theta_1,\theta_2)
\end{equation}
where
\begin{equation}
  i^{TQ}_\ell(\theta_1,\theta_2) = \frac{2\ell+1}{4\upi}
  \sqrt{\frac{(\ell-2)!}{(\ell+2)!}} \left[ \tilde\imath^{TQ}_\ell(\cos\theta_1)
    - \tilde\imath^{TQ}_\ell(\cos\theta_2) \right] .
\end{equation}
This vector may be precomputed for rapid evaluation of
$\sTQ(\theta_1,\theta_2)$.

\bsp

\label{lastpage}

\end{document}